\documentclass{PoS}

\title{Higher Spin Gravity and Exact Holography}

\ShortTitle{Higher Spin Gravity and Exact Holography}

\usepackage{amsmath,amssymb,amsfonts}
\usepackage{graphicx}

\def\half{\frac{1}{2}}

\newcommand{\bea}{\begin{eqnarray}}
\newcommand{\eea}{\end{eqnarray}}
\newcommand{\be}{\begin{equation}}
\newcommand{\ee}{\end{equation}}
\newcommand{\ba}{\begin{align}}
\newcommand{\ea}{\end{align}}

\newcommand{\w}{\mathcal{W}}

\newcommand{\h}{\mathcal{H}}

\DeclareMathOperator{\Tr}{Tr}

\author{\speaker{Kewang Jin} \\ %
        Institute for Theoretical Physics, ETH-Zurich, Switzerland \\
        E-mail: \email{jinke@itp.phys.ethz.ch}}

\abstract{In this talk, we present some direct evidences of the Higher Spin/Vector Model correspondence. There are two particular examples we would like to address on. The first example concerns a constructive approach of four dimensional higher spin theory from 3d $O(N)$ vector model based on a bi-local formulation. These bi-local fields are seen to give a bulk description of the higher spin theory with extra dimension and interactions. The second example is a similar AdS$_3$/CFT$_2$ duality put forward by Gaberdiel and Gopakumar. Specifically, we are interested in black hole solutions carrying nonzero higher spin charges. The partition function of the general $hs[\lambda]$ higher spin black hole is computed from the CFT side using the $\w_\infty[\lambda]$ symmetry and we found a perfect agreement with the gravity result.}

\FullConference{Proceedings of the Corfu Summer Institute 2012 \\
		 September 8-27, 2012 \\
		 Corfu, Greece}

\begin{document}

\section{Introduction}

The AdS/CFT correspondence \cite{Maldacena:1997re, Gubser:1998bc, Witten:1998qj} represents one of the major tools to understand Quantum Gravity. It is a special example of the Holographic principle \cite{'tHooft:1993gx, Susskind:1994vu} which states a quantum theory in $d+1$ dimensions with gravity can be described by a gauge theory on the boundary (with one lower dimension). Therefore, understanding the emergence of the extra dimension plays a central role to unravel the mechanism of the holographic principle. Among various examples of the AdS/CFT correspondence, the best studied one is of course the type IIB string theory on $AdS_5 \times S^5$ and its dual to ${\cal N}=4$ Supersymmetric Yang-Mills theory in four dimensions. However, this duality relates two highly non-trivial theories and crucially relies on supersymmetry. Furthermore, this is a strong-weak duality which makes the correspondence hard to prove analytically. Therefore it is very important and meaningful to find some simplified model of the AdS/CFT correspondence where both sides of duality may be exactly solvable. This will provide some direct evidences and definite understanding of the AdS/CFT correspondence.

The Higher Spin/Vector Model correspondence serves such an example. On the field theory side, it is the simplest $N$-component (free) vector model. In the large $N$ limit, a special kind of higher spin AdS gravity developed by Vasiliev and collaborators \cite{Vasiliev:1990en, Vasiliev:1992av} emerges. As pointed out by Klebanov and Polyakov \cite{Klebanov:2002ja}, the singlet sector of the $O(N)$ vector model is dual to Vasiliev higher spin gravity with minimal coupling to the scalar field (see also \cite{Sezgin:2002rt, Sezgin:2003pt}). The three-dimensional $O(N)$ model has two fixed points: a free (UV) fixed point and an interacting (IR) fixed point. To be more precise, these two fixed points are dual to two boundary conditions of the bulk scalar field.

On the gravity side, the higher spin theory of Vasiliev describes the interaction of an infinite tower of massless higher spin fields with gravity (and other lower spin matter fields). In four and higher dimensions, a simple action principle still remains to be found (see however \cite{Boulanger:2011dd} for an action principle of the ``extended'' Vasiliev's system), instead the Vasiliev theory is described in terms of nonlinear (and nonlocal) equations of motion. The higher spin theory is itself a gauge theory with a large class of higher spin gauge symmetry; and the gauge fixing of the full nonlinear theory is far from obvious. Without knowing an explicit action, the calculation of correlation functions through the equations of motion is highly technical. The process usually involves partially gauge fixing combined with partially solving the equations. An impressive comparison of three-point functions was made by Giombi and Yin \cite{Giombi:2009wh, Giombi:2010vg} who were able to show the two fixed points of the $O(N)$ vector model are dual to two boundary conditions of the bulk scalar field as conjectured by Klebanov and Polyakov  \cite{Klebanov:2002ja}.

We have in \cite{Das:2003vw, Koch:2010cy} formulated a constructive approach of bulk AdS higher spin gravity in terms of bi-local fields of the $O(N)$ model. These bi-local fields describe (over)-completely the singlet sector of the $O(N)$ model and leads to a nonlinear, interacting theory (with $1/N$ as the coupling constant) which was seen to possess all the properties of the dual AdS theory. The interactions are present for both the free and critical fixed points. More importantly, the bi-local formulation automatically reproduces arbitrary-point correlation functions and provides a construction of higher spin theory (in various gauges \cite{Jevicki:2011ss}) based on the CFT. The main idea of the construction is to study how the fields transform under the symmetry and compare directly the generators. A canonical transformation was found between the $SO(2,3)$ generators and the conformal generators in 3d, based on which an integral transformation was then found to map the bi-local fields to the higher spin fields with the extra AdS dimension. This map is one-to-one and provides direct evidence and understanding of the emergent AdS spacetime.

The second part of this talk concerns the duality between the AdS$_3$ higher spin theory (coupled to matter \cite{Prokushkin:1998bq}) and a 2d minimal model proposed by Gaberdiel and Gopakumar \cite{Gaberdiel:2010pz}. To be more precise, the 3d massless higher spin theory coupled with two massive scalars is dual to the 2d ${\cal W}_N$ minimal model in the 't Hooft limit (defined below). The $\w_N$ minimal model can be constructed using the WZW coset
\begin{equation}
\frac{su(N)_k \oplus su(N)_1}{su(N)_{k+1}}
\end{equation}
where the central charge at finite $N,k$ reads
\begin{equation}
c=(N-1) \left[ 1 -\frac{N(N+1)}{(N+k)(N+k+1)} \right] \ .
\end{equation}
In the 't Hooft limit defined by taking $N,k \to \infty$ and keeping the 't Hooft coupling constant fixed
\begin{equation}
0 \le \lambda \equiv \frac{N}{N+k} \le 1 \ ,
\label{tHooft}
\end{equation}
the central charge scales as $c \sim N(1-\lambda^2)$. Therefore, these models are vector-like (and the $\lambda$ parameter also fixes the mass of the AdS scalar field). The primaries of the $\w_N$ minimal model are labelled by $(\Lambda_+;\Lambda_-)$ where $\Lambda_+$ and $\Lambda_-$ are respectively the representations of $su(N)_k$ and $su(N)_{k+1}$. The representation of $su(N)_1$ is uniquely determined by the selection rule \cite{Bouwknegt:1992wg}.

Three dimensional higher spin theories are much simpler to study (than higher dimensional theories) because it is consistent to truncate the infinite tower of higher spin fields to a finite set of spins with maximal spin $N$, i.e. the higher spin theory with the spin content $s=2,3,...N$ is closed. More importantly, this theory has a Chern-Simons formulation with the gauge group $SL(N,\mathbb{R}) \times SL(N,\mathbb{R})$, i.e. the action is given by \cite{Blencowe:1988gj}
\begin{equation}
S_{HS} = S_{CS}[A]-S_{CS}[\bar{A}]
\end{equation}
where
\begin{equation}
S_{CS}[A] = \frac{k_{CS}}{4 \pi} \int {\rm Tr} (A \wedge d A + \frac{2}{3} A \wedge A \wedge A) \ ,
\end{equation}
and similarly for $S_{CS}[\bar{A}]$. The Chern-Simons level is related to the AdS radius by
\begin{equation}
k_{CS} = \frac{\ell}{4 G_N} \ ,
\end{equation}
where $G_N$ is Newton's constant. The Chern-Simons formulation can be extended to the infinite spin case,\footnote{Actually, the original derivation of \cite{Blencowe:1988gj} is for the infinite algebra $hs(1,1)=hs[\frac{1}{2}]$.} in which the relevant Lie algebra is the one-parameter family of higher spin algebra $hs[\lambda]$.\footnote{The $\lambda$ parameter is the same as the 't Hooft coupling constant (\ref{tHooft}).} It is an infinite dimensional algebra which can be realized as a quotient of the universal enveloping algebra of $sl(2)$ by a proper ideal \cite{Gaberdiel:2011wb}
\begin{equation}
hs[\lambda] \oplus \mathbb{C} = \frac{U(sl(2))}{\langle C_2 - \mu \mathbf{1} \rangle} \ ,
\end{equation}
where $C_2$ is the quadratic Casimir of $sl(2)$ and $\mu = \frac{1}{4} (\lambda^2 -1)$. The vector corresponding to $\mathbb{C}$ is the identity generator $\mathbf{1}$ of the universal enveloping algebra.  Setting $\lambda = N \ge 2$, an ideal $\chi_N$ consisting all the higher spin generators with $s>N$ appears. Quotienting out this ideal, one obtains the $sl(N)$ algebra
\begin{equation}
hs[\lambda = N]/\chi_N \cong sl(N) \ .
\label{ideal}
\end{equation}

Using the Chern-Simons formulation of higher spin gravity and following the Brown $\&$ Henneaux analysis for pure gravity \cite{Brown:1986nw}, the asympotitic symmetry of $SL(N)$ gravity is the $\w_N$ algebra \cite{Henneaux:2010xg, Campoleoni:2010zq, Campoleoni:2011hg}. Similarly, the asymptotic symmetry of the $hs[\lambda]$ gravity is the one-parameter family of $\w$-algebra denoted as $\w_\infty[\lambda]$ \cite{Gaberdiel:2011wb}. All these classical analyses lead to the same central charge of the $\w$-algebra (as the Virasoro algebra for pure gravity)
\begin{equation}
c=\frac{3 \ell}{2 G_N} = 6 k_{CS} \ .
\label{levelc}
\end{equation}
Similar to (\ref{ideal}), setting $\lambda=N$ and quotienting out an ideal, we obtain
\begin{equation}
\w_\infty [\lambda = N]/\chi_N \cong \w_N \ .
\end{equation}

However, the 't Hooft coupling constant defined in (\ref{tHooft}) is between 0 and 1; it is not obvious the asymptotic symmetry algebra $\w_\infty[\mu]$ with $\mu=\lambda$ is the same as the symmetry algebra of the coset model $\w_\infty[\mu]$ with $\mu=N$. This was clarified in \cite{Gaberdiel:2012ku} by studying the quantum algebra (with finite values of $N$ and $c$). Specifically, a triality isomorphism of the $\w$-algebra (at fixed $c$) was discovered
\begin{equation}
\w_\infty[N] \cong \w_\infty[\frac{N}{N+k}] \cong \w_\infty[-\frac{N}{N+k+1}] \ ,
\end{equation}
which explains the agreement of the symmetry algebra.\footnote{The last isomorphism, in the 't Hooft limit, leads to $\w_\infty[\lambda] \sim \w_\infty[-\lambda]$. This can be easily seen from the commutation relations of $\w_\infty[\lambda]$ algebra \cite{Gaberdiel:2012yb} that the structure constants only depend on $\lambda^2$. As a matter of fact, for the bulk higher spin theory, the analytic continuation of $hs[\lambda]$ gravity to $sl(N)$ gravity is achieved by $\lambda=-N$ \cite{Gaberdiel:2012ku, Perlmutter:2012ds}.} The quantum analysis also reveals that one of the scalar fields in the 't Hooft limit (which duals to the $(0;{\rm f})$ representation of the CFT) has a non-perturbative origin. This suggests the perturbative Vasiliev theory is not complete at the quantum level (various non-perturbative excitations must be added) \cite{Chang:2011mz, Papadodimas:2011pf, Gaberdiel:2012ku}. Besides the agreement of the symmetry algebra, the Gaberdiel-Gopakumar conjecture is also supported by the matching of partition functions \cite{Gaberdiel:2011zw} and certain correlation functions \cite{Chang:2011mz, Papadodimas:2011pf, Ahn:2011by, Ammon:2011ua, Chang:2011vka, Hijano:2013fja}, etc.

\section{Direct Construction of Higher Spin Gravity from the $O(N)$ Model}

Starting with the AdS$_4$/CFT$_3$ case, the Lagrangian of the 3d $O(N)$ model is given by
\begin{equation}
L=\int d^3 x \left( \frac{1}{2} (\partial_\mu \phi^a)(\partial^\mu \phi^a)+ \frac{g}{4 \, N}(\phi \cdot \phi)^2 \right), \qquad a=1,...,N.
\end{equation}
As mentioned before, this model has two fixed points: the UV fixed point with zero coupling constant ($g=0$) and the IR fixed point with non-zero coupling constant. On the bulk side, the mass of the scalar field is $m^2=-2$ which gives two possible boundary conditions with scaling dimensions $\Delta_-=1$ and $\Delta_+=2$. The Klebanov-Polyakov conjecture states that the $\Delta_-=1$ boundary condition is dual to the free $O(N)$ model, whereas the $\Delta_+=2$ boundary condition duals to the critical $O(N)$ model. These two fixed points are related by a Legendre transformation \cite{Klebanov:1999tb, Petkou:2003zz} (see also \cite{Witten:2001ua, Gubser:2002vv, Hartman:2006dy}). Therefore, the duality between Vasiliev's theory and the critical $O(N)$ model follows, order by order in $1/N$, from the duality with free $O(N)$ model \cite{Giombi:2011ya}. While for collective field theory (reviewed in the next subsection), the critical $O(N)$ model can be treated on equal footing as the free $O(N)$ model. In the following, we will focus on the free $O(N)$ model whereas trying to keep some formulae valid also for the critical model.

\subsection{Collective field theory of the $O(N)$ model}
\label{sec:collective}

The collective field theory we employ here was developed by Jevicki and Sakita \cite{Jevicki:1979mb} for large-$N$ quantum field theory. The main idea is to reformulate the theory using the invariant (under the symmetry group) collective fields. After a non-trivial change of variables, a collective action (for the Euclidean case) or Hamiltonian (for the Minkowski case) can be derived in terms of the collective fields (and their conjugate momenta). A subsequent $1/N$ expansion can be carried out in a rather straightforward way. This method had great success in applying to the $U(N)$ matrix model \cite{Jevicki:1979mb}, which leads to a field theoretic formulation of $D=1$ string as a massless scalar field in two dimensions \cite{Das:1990kaa}. There, the emergence of extra dimension comes from the large-$N$ color index of the matrix model.

The constructive approach for the AdS$_4$/CFT$_3$ duality formulated in \cite{Das:2003vw} is based on the $O(N)$ invariant bi-local fields
\begin{eqnarray}
\Phi(x,y) \equiv \sum_{a=1}^N \phi^a(x) \cdot \phi^a(y)
\end{eqnarray}
which close under the Schwinger-Dyson equations in the large $N$ limit. Through a series of chain rules of the type
\begin{equation}
\frac{\partial}{\partial \phi^a(x)} = \int dy \int dz \, \frac{\partial \Phi(y,z)}{\partial \phi^a(x)} \frac{\partial}{\partial \Phi(y,z)}
= \int dy \, \phi^a (y) \left[ \frac{\partial}{\partial \Phi(y,x)} + \frac{\partial}{\partial \Phi(x,y)} \right] \ ,
\label{chain}
\end{equation}
the collective action which evaluates the complete $O(N)$ invariant partition function can be derived as \cite{Jevicki:1980zg}
\begin{equation}
Z=\int [\prod_a d\phi^a(x)] e^{-S[\phi]}=\int [d\Phi(x,y)] \mu(\Phi) e^{-S_c[\Phi]}
\end{equation}
where the measure is given by $\mu(\Phi)=(\det \Phi)^{V_x V_p}$ with $V_x=L^3$ the volume of space and $V_p=\Lambda^3$ the volume of momentum space (where $\Lambda$ is the momentum cutoff). Explicitly the collective action can be computed as in \cite{Jevicki:1980zg, Das:2003vw} to be
\begin{equation}
S_c[\Phi] = \int dx \left[-\frac{1}{2} \lim_{y \to x} \partial_x^2 \Phi(x,y) + \frac{g}{4 \, N} \Phi^2 (x,x) \right]-\frac{N}{2} \int dx \, \ln \Phi(x,x) \ ,
\label{actionEucl}
\end{equation}
where the first term is a direct rewriting of the original action in terms of bi-local fields; while the second interaction term $\ln \Phi$ arises from the Jacobian of the change of variables
\begin{equation}
\int \prod_a d\phi^a (x) = \int d\Phi(x,y) J[\Phi] \ .
\end{equation}
We stress that the Jacobian here gives both the measure $\mu(\Phi)$ and the interaction term $\ln \Phi$.

\begin{figure}
\centering
\includegraphics[width=.7\textwidth]{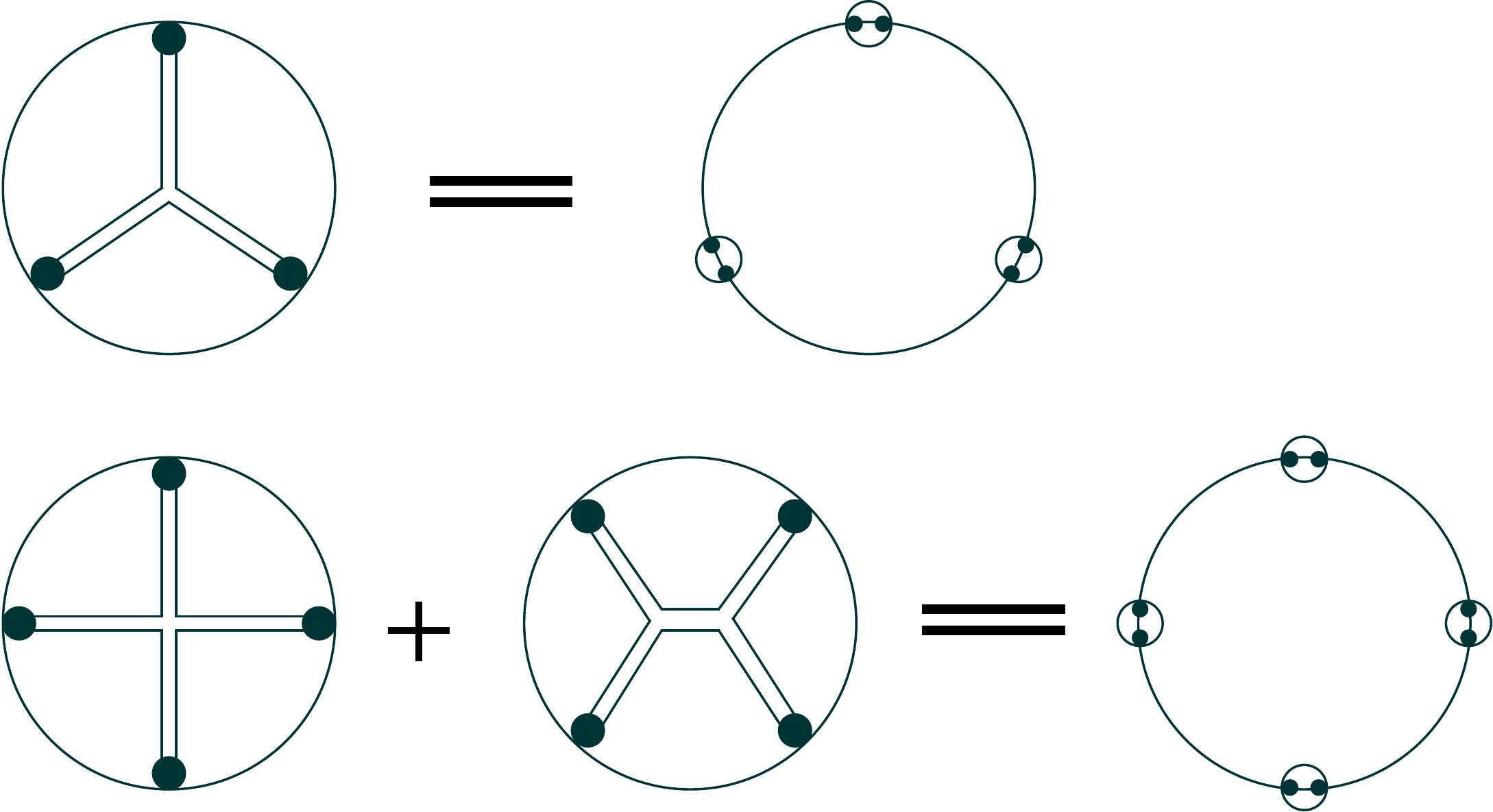}
\caption{Illustration of three- and four-point collective field diagrams.}
\label{fourdiagram}
\end{figure}

The collective action (\ref{actionEucl}) is nonlinear, with $1/N$ appearing as the expansion parameter. The perturbative expansion in this bi-local theory proceeds in the standard way. The nonlinear equation of motion specified by $S_c$ gives the background
\begin{equation}
\left. \Phi \frac{\partial S_c}{\partial \Phi} \right|_{\Phi = \Phi_0} = 0 \Longrightarrow - \partial_x^2 \Phi_0 (x,y) + \frac{g}{N} \Phi_0^2 (x,y) - 2 N \delta(x-y) = 0 \ .
\end{equation}
Expanding around the background $\Phi=\Phi_0-\frac{1}{\sqrt{N}}\eta$ gives an infinite sequence of interaction vertices \cite{deMelloKoch:1996mj} 
\begin{eqnarray}
S_c[\Phi] = S_c[\Phi_0]+{\rm Tr}\left[ \frac{1}{4} \Phi_0^{-1} \eta \Phi_0^{-1} \eta +\frac{g}{4 \, N^2}\eta^2+\sum_{n \ge 3} \frac{1}{2n} N^{1-\frac{n}{2}} (\Phi_0^{-1} \eta)^n \right] \ ,
\end{eqnarray}
where the trace is defined to be ${\rm Tr}[A(x,y) B(y,z)] = \int dx \int dy \, A(x,y) B(y,x)$. The nonlinearities built into $S_c$ are precisely such that all invariant $n$-point correlators of the $O(N)$ singlet fields
\begin{equation} 
\langle \Phi(x_1,y_1) \cdots \Phi(x_n,y_n) \rangle = \langle \phi(x_1) \cdot \phi(y_1) \cdots \phi(x_n) \cdot \phi(y_n) \rangle
\end{equation}
are reproduced through the Witten diagrams with $1/N$ vertices as shown in Figure \ref{fourdiagram}. We stress that this nonlinear structure is there for both the interacting and the free ($g=0$) fixed points.

This bi-local theory is expected to represent a covariant-type gauge fixing of Vasiliev's gauge invariant theory. A large number of degrees of freedom are removed in fixing a gauge and this happens in higher spin gravity too. In section \ref{sec:symmetric}, we will make a connection of this covariant bi-local theory to a symmetric gauge of the Vasiliev theory. We can show this reduced bulk system (after further solving some components of the equations of motion) has the same dimensionality as the covariant bi-local theory.

A one-to-one relationship between bi-local fields and AdS higher spin fields are demonstrated in a physical gauge with a single time \cite{Koch:2010cy}. The existence of such a gauge is not a priori obvious (because this is a reduction of two-time physics to a single-time physics where nontrivial issues of unitarity should be addressed). In \cite{Jevicki:2011ss}, we have shown such a gauge fixing and a discussion of the {\it collective dipole} underlying the collective construction was given.

The single-time formulation involves the equal-time bi-local fields
\begin{equation}
\Psi(t,\vec{x},\vec{y})=\sum_a \phi^a(t,\vec{x}) \phi^a(t,\vec{y})
\label{bilocal}
\end{equation}
which are local in time but bi-local in $d-1$ spatial dimensions. These observables (collective fields) are characterized by the fact that they represent a complete set of $O(N)$ invariant canonical variables. Writing the conjugate momenta as
\begin{equation}
\Pi(\vec{x},\vec{y}) = -i \frac{\partial}{\partial \Psi(\vec{y},\vec{x})} \ ,
\end{equation}
the collective Hamiltonian is of the form \cite{Jevicki:1980mj, Jevicki:1983hb}
\begin{equation}
H=2 \int d\vec{x} \, d\vec{y} \, d\vec{z} \, \, \Pi(\vec{x},\vec{y}) \Psi(\vec{y},\vec{z}) \Pi(\vec{z},\vec{x})+ V[\Psi]
\label{HamMink}
\end{equation}
where the potential term reads
\begin{equation}
V[\Psi] = \int d\vec{x} \left[ -\frac{1}{2} \lim_{\vec{y} \to \vec{x}} \nabla_{\vec{x}}^2 \Psi(\vec{x},\vec{y}) + \frac{g}{4 \, N} \Psi^2 (\vec{x},\vec{x}) \right]+\frac{N^2}{8} \int d\vec{x} \, \, \Psi^{-1}(\vec{x},\vec{x}) \ .
\label{lastMink}
\end{equation}
Similar to the Euclidean case, the first term is a direct rewriting of the original potential term in terms of the bi-local fields. The appearance of the last term needs an explanation. In general, a direct reformulation of the original Hamiltonian in terms of the bi-local fields (after applying the chain rules (\ref{chain})) is not hermitian. There exists a similarity transformation after which the final effective Hamiltonian is hermitian. The last term in (\ref{lastMink}) arises precisely from such a similarity transformation and this is in some sense analogous to the Jacobian of the Euclidean case.

The Hamiltonian (\ref{HamMink}) again has a natural $1/N$ expansion, after a background shift
\begin{equation}
\Psi=\Psi_0+\frac{1}{\sqrt{N}} \eta \ , \qquad \Pi=\sqrt{N} \pi \ ,
\end{equation}
where $\Psi_0$ is the saddle point of $V[\Psi]$, the quadratic Hamiltonian reads
\begin{eqnarray}
H^{(2)} = {\rm Tr} ( \pi \Psi_0 \pi ) + \frac{1}{8} {\rm Tr} \left( \Psi_0^{-1} \eta \Psi_0^{-1} \eta \Psi_0^{-1} \right)
\end{eqnarray}
where we have set the coupling constant $g=0$ for simplicity.\footnote{There is an overall constant $N$ we omit here (and subsequently for all the higher vertices), which states the coupling constant of the collective field theory is $G_N = 1/N$.} Fourier transforming the fluctuations $\eta,\pi$ as well as the background field $\Psi_0$
\begin{eqnarray}
\Psi^0_{\vec{x}\vec{y}} &=& \int d\vec{k} \, e^{i \vec{k} \cdot (\vec{x}-\vec{y})} \psi_{\vec{k}}^0 \ , \qquad \psi_{\vec{k}}^0 = \frac{1}{2 \sqrt{\vec{k}^2}} \\
\eta_{\vec{x}\vec{y}} &=& \int d\vec{k}_1 d\vec{k}_2 e^{-i \vec{k}_1 \cdot \vec{x} + i \vec{k}_2 \cdot \vec{y}} \eta_{\vec{k}_1 \vec{k}_2} \ , \\
\pi_{\vec{x}\vec{y}} &=& \int d\vec{k}_1 d\vec{k}_2 e^{+i \vec{k}_1 \cdot \vec{x} - i \vec{k}_2 \cdot \vec{y}} \pi_{\vec{k}_1 \vec{k}_2} \ ,
\end{eqnarray}
one finds the quadratic Hamiltonian in momentum space
\begin{equation}
H^{(2)}=\frac{1}{2} \int d \vec{k}_1 d \vec{k}_2 \; \pi_{\vec{k}_1 \vec{k}_2} \pi_{\vec{k}_1 \vec{k}_2}+\frac{1}{8} \int d \vec{k}_1 d \vec{k}_2 \; \eta_{\vec{k}_1 \vec{k}_2}
\left( \psi_{\vec{k}_1}^{0\,\,-1}+\psi_{\vec{k}_2}^{0\,\,-1} \right)^2\eta_{\vec{k}_1 \vec{k}_2}   
\end{equation}
which correctly describes the (singlet) spectrum of the $O(N)$ theory with $\omega_{\vec{k}_1 \vec{k}_2} = \sqrt{\vec{k}_1^2} + \sqrt{\vec{k}_2^2}$. Higher vertices representing $1/N$ interactions can be found similarly, in particular, the cubic and quartic interactions are given explicitly as
\begin{eqnarray}
H^{(3)}&=&\frac{2}{\sqrt{N}}{\rm Tr}(\pi \eta \pi)-\frac{1}{8\sqrt{N}}{\rm Tr} \left( \Psi_0^{-1} \eta \Psi_0^{-1} \eta \Psi_0^{-1} \eta \Psi_0^{-1} \right) \ , \label{cubic} \\
H^{(4)}&=&\frac{1}{8N}{\rm Tr} \left( \Psi_0^{-1} \eta \Psi_0^{-1} \eta \Psi_0^{-1} \eta \Psi_0^{-1} \eta \Psi_0^{-1} \right) \ . \label{quartic}
\end{eqnarray}
We note that the form of these vertices is the same for both the free (UV) and the interacting (IR) conformal theories. The only difference is induced by different background shifts $\Psi_0$ in these two cases.

\subsection{Scattering matrix}

For the UV fixed point of the vector model, there exists an infinite sequence of exactly conserved higher spin currents. Consequently one has a higher symmetry with infinite number of generators. In such a theory, the Coleman-Mandula theorem would imply the $S$ matrix should be 1. In general, there is a question regarding the existence of an $S$ matrix in CFT (and also in AdS gravity). Maldacena and Zhiboedov \cite{Maldacena:2011jn} considered the implication of this theorem on correlation functions. In particular, they have shown the correlation functions $\langle {\cal O}_1 {\cal O}_2 \cdots {\cal O}_n \rangle$ of a CFT in the presence of higher spin symmetry can be written as that of free fields (free bosons or free fermions). However, the correlators themselves are nonzero and nontrivial for all $n$.

We consider the scattering of ``collective dipoles'' just introduced and calculate the collective $S$-matrix defined by the LSZ-type reduction formula
\begin{eqnarray}
S=\lim \prod_i (E_i^2-(\vert \vec{k}_i \vert+\vert \vec{k}_{i'} \vert )^2) \langle \tilde{\Psi}(E_1,\vec{k}_1,\vec{k}_{1'}) \tilde{\Psi}(E_2,\vec{k}_2,\vec{k}_{2'}) \cdots \rangle
\label{Sdefintion}
\end{eqnarray}
where $\tilde{\Psi}$ is the energy-momentum transform of the bi-local field (\ref{bilocal}). The limit implies the on-shell specification for the energies of the dipoles: $E^2-(\vert \vec{k}_1 \vert+\vert \vec{k}_2 \vert)^2=0$.

Our evaluation of the $S$-matrix proceeds as follows. Using the time-like quantization we will evaluate the 3 and 4-point scattering amplitude corresponding to the associated Witten diagrams. In momentum space, in terms of the bi-local fields
\begin{eqnarray}
\eta(t;\vec{x}_1,\vec{x}_2) &=& \int d\vec{k}_1 d\vec{k}_2\frac{1}{\sqrt{2\omega_{k_1}\omega_{k_2}}} \left(e^{+i(\vec{k}_1 \cdot \vec{x}_1+\vec{k}_2 \cdot \vec{x}_2)}\alpha_{\vec{k}_1\vec{k}_2}+h.c. \right) \\
\pi(t;\vec{x}_1,\vec{x}_2) &=& i \int d\vec{k}_1 d\vec{k}_2 \sqrt{\frac{\omega_{k_1}\omega_{k_2}}{2}} \left(e^{-i(\vec{k}_1 \cdot \vec{x}_1+\vec{k}_2 \cdot \vec{x}_2)}\alpha^\dagger_{\vec{k}_1\vec{k}_2}-h.c. \right)
\end{eqnarray}
the cubic (\ref{cubic}) and quartic (\ref{quartic}) interactions take the form
\begin{eqnarray}
H^{(3)} &=& \frac{\sqrt{2}}{\sqrt{N}}\int \prod_{i=1}^3 d\vec{k}_i \Bigl[-\frac{\omega_{k_1 k_2 k_3}}{3} \alpha_{\vec{k}_1 \vec{k}_2}\alpha_{-\vec{k}_2 \vec{k}_3}\alpha_{-\vec{k}_3-\vec{k}_1}+\omega_{k_2} \alpha_{\vec{k}_1 \vec{k}_2}\alpha_{-\vec{k}_2 \vec{k}_3}\alpha^\dagger_{\vec{k}_3 \vec{k}_1}+h.c. \Bigr] 
\label{cubicosc}
\end{eqnarray}
\begin{eqnarray}
H^{(4)} &=& \frac{1}{N} \int \prod_{i=1}^4 d\vec{k}_i \; \frac{\omega_{k_1 k_2 k_3 k_4}}{4} \Bigl[\alpha_{\vec{k}_1 \vec{k}_2}\alpha_{-\vec{k}_2 \vec{k}_3}\alpha_{-\vec{k}_3 \vec{k}_4}\alpha_{-\vec{k}_4-\vec{k}_1}+4\alpha_{\vec{k}_1 \vec{k}_2}\alpha_{-\vec{k}_2 \vec{k}_3}\alpha_{-\vec{k}_3 \vec{k}_4}\alpha^\dagger_{\vec{k}_4 \vec{k}_1}+h.c. \cr
&& \qquad\qquad\qquad\qquad+4\alpha_{\vec{k}_1 \vec{k}_2}\alpha_{-\vec{k}_2 \vec{k}_3}\alpha^\dagger_{\vec{k}_3 \vec{k}_4}\alpha^\dagger_{-\vec{k}_4 \vec{k}_1}+2\alpha_{\vec{k}_1 \vec{k}_2}\alpha^\dagger_{\vec{k}_2 \vec{k}_3}\alpha_{\vec{k}_3 \vec{k}_4}\alpha^\dagger_{\vec{k}_4 \vec{k}_1} \Bigr] 
\label{quarticosc}
\end{eqnarray}
where we have used the notation $\omega_{k_1 k_2 \cdots k_i} \equiv \omega_{k_1} + \omega_{k_2} + \cdots + \omega_{k_i}$ and $h.c.$ means the hermitian conjugate of {\it only} the terms ahead of it.

\begin{figure}[htp]
\centering
\begin{tabular}{ccc}
\includegraphics[width=.23\textwidth]{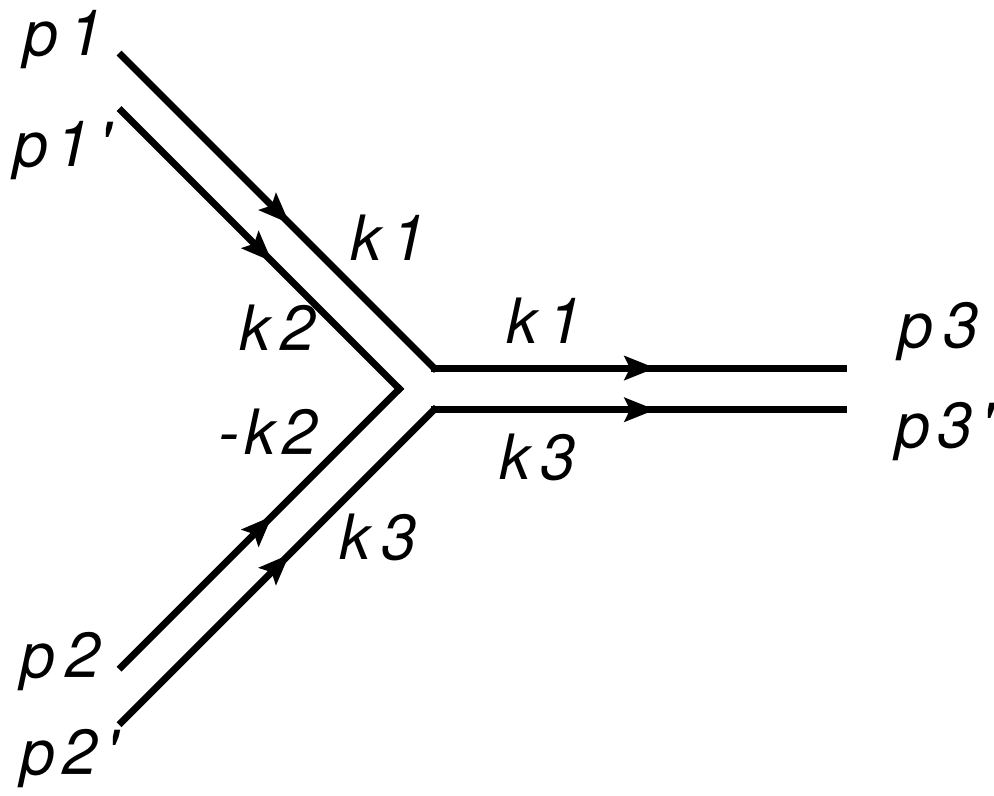} &
\includegraphics[width=.3\textwidth]{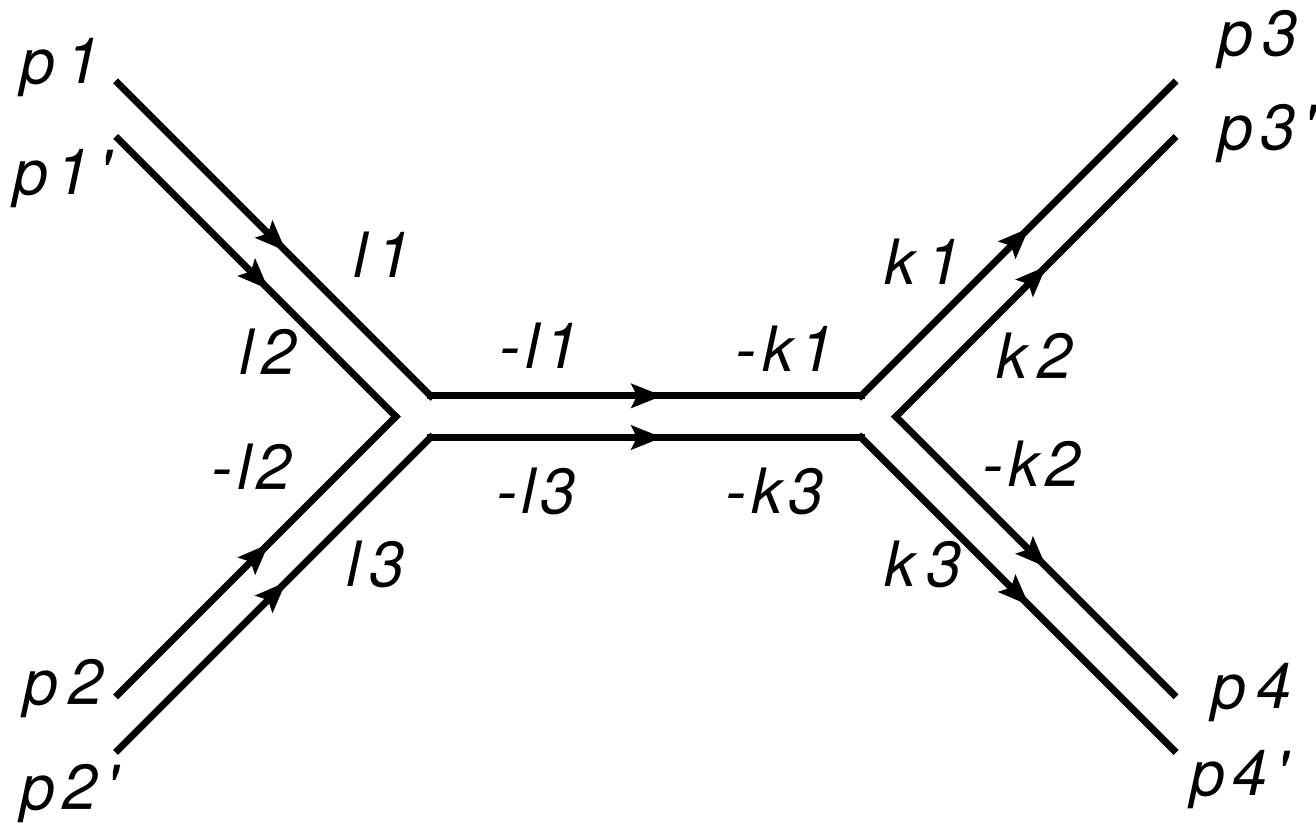} &
\includegraphics[width=.21\textwidth]{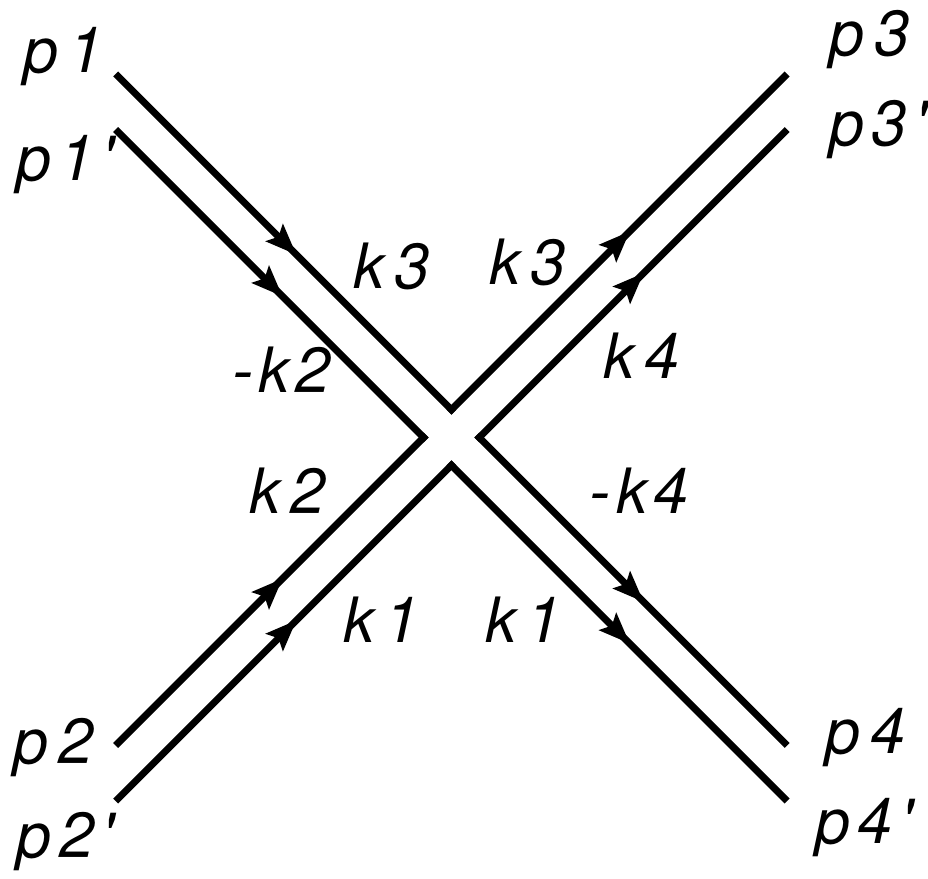} \cr
(a) & (b) & (c)
\end{tabular}
\caption{The scattering of three and four collective dipoles.}
\label{scattering}
\end{figure}

For the three-dipole scattering ($1+2 \to 3$), the amplitude is given by
\begin{equation}
\langle 0 \vert \alpha_{\vec{p}_{3} \vec{p}_{3'}} \,
T \exp \left[ -i \int_{-\infty}^\infty dt \, H^{(3)} (t) \right]
\alpha^\dagger_{\vec{p}_{2} \vec{p}_{2'}} \alpha^\dagger_{\vec{p}_{1} \vec{p}_{1'}} \vert 0 \rangle
\end{equation}
where $T$ means {\it time-ordered}. Graphically, this corresponds to evaluating the Feynman diagram as shown in Figure \ref{scattering}a. The evaluation involves using the bi-local propagator symmetrized over the momenta
\begin{align}
\langle 0\vert T \alpha_{\vec{p}_{1} \vec{p}_{1'}}(t_1)
\alpha^\dagger_{\vec{p}_{2} \vec{p}_{2'}}(t_2)\vert 0 \rangle
=\int dE\frac{ie^{-iE(t_1-t_2)}}{E-\omega_{p_1}-\omega_{p_{1'}}}
& \frac{1}{2}[\delta(\vec{p}_1-\vec{p}_2)\delta(\vec{p}_{1'}-\vec{p}_{2'}) \cr
& +\delta(\vec{p}_1-\vec{p}_{2'})\delta(\vec{p}_{1'}-\vec{p}_2)] \ .
\end{align}
The on-shell three-point scattering amplitude is obtained by amputating the leg poles and putting the external states on-shell which gives the final result (see \cite{deMelloKoch:2012vc} for more details)
\begin{align}
S(1+2\rightarrow 3)=&-\frac{\sqrt{2}}{8\sqrt{N}} (E_1+E_2-E_3) \; \delta(E_1+E_2-E_3) \cr
&\times [ \delta(\vec{p}_1-\vec{p}_3) \delta(\vec{p}_{2'}-\vec{p}_{3'}) \delta(\vec{p}_{1'}+\vec{p}_2) + \text{7 more terms} ] \ .
\label{finalthree}
\end{align}
The seven more terms in the end are due to the symmetrization over $(1 \leftrightarrow 1'), (2 \leftrightarrow 2'), (3 \leftrightarrow 3')$. The final result (\ref{finalthree}) is of the form $x \delta (x)$, therefore $S_3=0$ follows.

Next for the four-dipole scattering ($1+2 \to 3+4$), the calculation is similar. The scattering amplitude is given by
\begin{equation}
\langle 0\vert \alpha_{\vec{p}_{3}\vec{p}_{3'}}\alpha_{\vec{p}_{4}\vec{p}_{4'}} T \exp \left[
-i\int^\infty_{-\infty} dt \, \left( H^{(3)} (t) + H^{(4)} (t) \right) \right]
\alpha^\dagger_{\vec{p}_{1} \vec{p}_{1'}}\alpha^\dagger_{\vec{p}_{2} \vec{p}_{2'}}\vert 0 \rangle
\end{equation}
where $H^{(4)}$ is explicitly given in (\ref{quarticosc}). Summing all the $s,t,u$-channel diagrams in Figure \ref{scattering}b and the cross-shaped diagrams in Figure \ref{scattering}c, the final result is \cite{deMelloKoch:2012vc}
\begin{eqnarray}
& S(1+2\rightarrow 3+4)= \frac{i}{16 N} (E_1+E_2-E_3-E_4) \delta(E_1+E_2-E_3-E_4)  \cr
& \times \bigl[ \delta(\vec{p}_1-\vec{p}_3) \delta(\vec{p}_{1'}+\vec{p}_2) \delta( \vec{p}_{2'}-\vec{p}_{4'}) \delta (\vec{p}_{3'}+\vec{p}_4) + \text{15 more terms} \cr
& \quad + \delta(\vec{p}_2-\vec{p}_3) \delta(\vec{p}_1+\vec{p}_{2'}) \delta( \vec{p}_{1'}-\vec{p}_{4'}) \delta (\vec{p}_{3'}+\vec{p}_4) + \text{15 more terms} \bigr] \ ,
\label{finalzero}
\end{eqnarray}
which implies $S_4=0$. We need to stress that the vanishing of the $S$-matrix is due to some genuine cancellations between different Feynman diagrams which are {\it not} individually zero.

It is clear that the direct evaluation can be continued to higher points with the conjectured result $S_{n \ge 5}=0$. One can describe the nonlinear collective field theory in the following way: its nonlinearity, and higher point vertices are precisely such that they reproduce the boundary correlators through the bi-local (Witten) diagrams. The sum of these diagrams however give vanishing results in the on-shell evaluation as described above. $S=1$ implies triviality. Consequently, using the equivalence theorem, the nonlinearities built into the $O(N)$ model can be transformed away under field redefinitions. We have in \cite{deMelloKoch:2012vc} described such a field transformation that leads to a quadratic Hamiltonian.

\subsection{One-to-one mapping in the light-cone gauge}

Our goal is to demonstrate that the collective field contains all the necessary information and is in a one-to-one map with the physical fields of the higher-spin theory in AdS$_4$. For this comparison to be done it is advantageous to work in the light-cone gauge, where the physical degrees of freedom are most transparent. Our strategy is to compare directly the action of the conformal group of the $d=3$ field theory with that of the AdS$_4$ higher spin fields. In this direct comparison we will see as expected a very different set of spacetime variables and a different realization of $SO(2,3)$. The number of canonical variables however will be shown to be identical and one can search for a (canonical) transformation to establish a one-to-one relation between the two representations.

The conformal generators of the $O(N)$ model in the null-plane quantization ($x^+=t$) can be worked out as in \cite{Koch:2010cy} using Noether's theorem. Denoting the transverse coordinates using the index $i=1,2,...,d-2$ and the conjugate momenta of $(x_1^-,x_2^-,x_1^i,x_2^i)$ as
\begin{equation}
p_1^+ = \frac{\partial}{\partial x_1^-} \ , \qquad
p_2^+ = \frac{\partial}{\partial x_2^-} \ , \qquad
p_1^i = \frac{\partial}{\partial x_1^i} \ , \qquad
p_2^i = \frac{\partial}{\partial x_2^i} \ ,
\end{equation}
in the case of $d=3$, the 10 conformal generators (acting on the bi-local field $\Psi(t; x_1^-, x_2^-, x_1^i, x_2^i)$) are listed as follows
\begin{eqnarray}
P^-&=&p_1^-+p_2^-=-\Bigl(\frac{p_1^i p_1^i}{2p_1^+}+\frac{p_2^i p_2^i}{2p_2^+}\Bigr) \label{cft1} \\
P^+&=&p_1^++p_2^+ \label{cft2} \\
P^i&=&p_1^i+p_2^i \label{cft3} \\
M^{+-}&=&t P^--x_1^- p_1^+-x_2^- p_2^+ \label{cft4} \\
M^{+i}&=&t P^i-x_1^i p_1^+ -x_2^i p_2^+ \label{cft5} \\
M^{-i}&=&x_1^- p_1^i+x_2^- p^i_2+x^i_1 \frac{p^j_1 p_1^j}{2p_1^+}+x^i_2 \frac{p^j_2 p_2^j}{2p_2^+} \label{cft6} \\
D&=&t P^-+x_1^- p_1^+ + x_2^- p_2^+ + x^i_1 p_1^i +x^i_2 p_2^i+ 2d_\phi \label{cft7} \\
K^-&=&x^i_1 x_1^i \frac{p^j_1 p_1^j}{4p_1^+}+x^i_2 x_2^i \frac{p^j_2 p_2^j}{4p_2^+}
+x_1^-(x_1^- p_1^++x^i_1 p_1^i +d_\phi) \cr
&&+x_2^-(x_2^- p_2^++x^i_2 p_2^i+d_\phi) \label{cft8} \\
K^+&=&t^2 P^-+t(x^i_1 p_1^i+x^i_2 p_2^i+2d_\phi)
-\frac{1}{2} x^i_1 x_1^i p_1^+ -\frac{1}{2} x^i_2 x_2^i p_2^+ \label{cft9} \\
K^i&=&-t \Bigl(x_1^i \frac{p^j_1 p_1^j}{2p_1^+}+x^i_2 \frac{p^j_2 p_2^j}{2p_2^+}+x_1^- p^i_1+x_2^- p^i_2\Bigr)
-\frac{1}{2} x^j_1 x_1^j p_1^i-\frac{1}{2} x^j_2 x_2^j p_2^i \cr
&&+x_1^i(x_1^- p_1^++x_1^j p_1^j+d_\phi)
+x_2^i(x_2^- p_2^++x_2^j p_2^j+d_\phi) \label{cft10}
\end{eqnarray}
where $d_\phi = \frac{1}{2}$ is the scaling dimension of a single boson.

Turing to the AdS side, the gauge invariant equations of motion for free higher spin theory was first worked out by Fronsdal \cite{Fronsdal:1978vb}, which takes the form \cite{Mikhailov:2002bp}
\begin{eqnarray}
\nabla_\rho \nabla^\rho h_{\mu_1...\mu_s}-s\nabla_\rho \nabla_{\mu_1}h^\rho_{~\mu_2...\mu_s}+\frac{1}{2}s(s-1)\nabla_{\mu_1} \nabla_{\mu_2}h^\rho_{~\rho \mu_3...\mu_s} \cr
+2(s-1)(s+d-2)h_{\mu_1...\mu_s}=0
\label{HSeom}
\end{eqnarray}
where $\nabla$ is the covariant derivative in AdS space. This can also be derived from linearization of Vasiliev's full nonlinear theory. The higher spin fields $h_{\mu_1 ... \mu_s}$ satisfy the double traceless condition
\begin{equation}
g_{\mu_1 \mu_2} g_{\mu_3 \mu_4} h^{\mu_1 \mu_2 \mu_3 \mu_4 ... \mu_s} = 0
\end{equation}
which becomes important when $s \ge 4$. In order to fix the light-cone gauge, it is more convenient to use the tangent space tensor fields defined by
\begin{equation}
\h^{A_1 ... A_s} = e_{\mu_1}^{A_1} \cdots e_{\mu_s}^{A_s} \, h^{\mu_1 .... \mu_s}
\label{double}
\end{equation}
where $e_\mu^A = \frac{1}{z} \delta_\mu^A$ is the frame one-form. Furthermore, one can construct the Fock space vector using the creation and annihilation operators $\alpha^A,\bar{\alpha}^A$ as
\begin{equation}
\vert \h \rangle = \h^{A_1 ... A_s} \alpha_{A_1} \cdots \alpha_{A_s} \vert 0 \rangle \ , \qquad \bar{\alpha}^A \vert 0 \rangle = 0 \ ,
\end{equation}
where the commutators are $[\bar{\alpha}^A,\alpha^B] = \eta^{AB}$, $[\alpha^A, \alpha^B] = [\bar{\alpha}^A, \bar{\alpha}^B ] = 0$. The light-cone gauge is fixed by solving the following set of constraints \cite{Metsaev:1999ui}
\begin{eqnarray}
&& \bar{\alpha}^+ \vert \h \rangle = 0 \\
&& \alpha^I \bar{\alpha}^I \vert \h \rangle = s \vert \h \rangle \\
&& \bar{\alpha}^I \bar{\alpha}^I \vert \h \rangle = 0 \\
&& \bar{\alpha}^- \vert \h \rangle = \left( - \frac{\partial^I}{\partial^+} \bar{\alpha}^I + \frac{s+d-1}{\partial^+} \bar{\alpha}^z - \frac{2(\partial^+ - \alpha^+ \bar{\alpha}^z)}{\partial^+(\partial^+-2\alpha^+\bar{\alpha}^z)} \bar{\alpha}^z \right) \vert \h \rangle
\end{eqnarray}
where $I$ is the transverse coordinates including the extra AdS dimension, i.e. $x^I=(x^i,z)$. In the case of AdS$_4$ (where $d=3$), the Poincar\'{e} coordinates are denoted as $(x^+=t,x^-,x,z)$. Furthermore, we have defined a higher spin coordinate $\theta$ coming from the spin matrix of the $xz$ plane (see \cite{Koch:2010cy} for details). Therefore, the AdS$_4$ higher spin fields also live in a 5-dimensional coordinate space, i.e. $\h(t;x^-,x,z;\theta)$. Denoting the conjugate momenta of $(x^-,x,z,\theta)$ as $(p^+,p^x,p^z,p^\theta)$, the $SO(2,3)$ isometry generators written in the ``conformal'' form are \cite{Metsaev:1999ui}
\begin{eqnarray}
P^-&=&-\frac{p^x p^x+p^z p^z}{2p^+} \label{ads1} \\
P^+&=&p^+ \label{ads2} \\
P^x&=&p^x \label{ads3} \\
M^{+-}&=&t P^--x^- p^+ \label{ads4} \\
M^{+x}&=&t p^x-x p^+ \label{ads5} \\
M^{-x}&=&x^- p^x-x P^-+ \frac{p^\theta p^z}{p^+} \label{ads6} \\
D&=&t P^-+x^- p^++x p^x+z p^z+d_a \label{ads7} \\
K^-&=&-\frac{1}{2}(x^2+z^2) P^-+x^-(x^-p^++x p^x+z p^z+d_a) \cr
&&+\frac{1}{p^+}\bigl((x p^z-z p^x)p^\theta+(p^\theta)^2\bigr) \label{ads8} \\
K^+&=&t^2 P^-+t(x p^x+z p^z+d_a)-\frac{1}{2}(x^2+z^2)p^+, \label{ads9} \\
K^x&=&t(x P^--x^- p^x- \frac{p^\theta p^z}{p^+})+ \frac{1}{2}(x^2-z^2) p^x \cr
&&+x(x^- p^++z p^z+d_a)+z p^\theta \label{ads10}
\end{eqnarray}
where $d_a=1$ is the scaling dimension of the creation operator $\alpha$ in AdS$_4$.

Working at the classical level where we ignore the constants (which will receive quantum corrections from normal ordering), a canonical transformation relating the bi-local generators (\ref{cft1}-\ref{cft10}) with the isometry generators (\ref{ads1}-\ref{ads10}) was found in \cite{Koch:2010cy} to be
\begin{eqnarray}
x^-&=&\frac{x_1^- p_1^++x_2^- p_2^+}{p_1^++p_2^+} \\
p^+&=&p_1^+ + p_2^+ \\
x&=&\frac{x_1 p_1^+ + x_2 p_2^+ }{p_1^++p_2^+} \\
p^x&=&p_1+p_2 \\
z&=&\frac{(x_1-x_2)\sqrt{p_1^+ p_2^+} }{p_1^+ + p_2^+} \\
p^z&=&\sqrt{\frac{p_2^+}{p_1^+}} p_1-\sqrt{\frac{p_1^+}{p_2^+}} p_2 \\
\theta &=& 2\arctan\sqrt{\frac{p_2^+}{p_1^+}} \\
p^\theta&=&\sqrt{p_1^+ p_2^+}(x_1^--x_2^-)
+\frac{x_1-x_2}{2}\Bigl(\sqrt{\frac{p_2^+}{p_1^+}} p_1+\sqrt{\frac{p_1^+}{p_2^+}} p_2\Bigr) \ . 
\end{eqnarray}
This establishes at the quadratic level the bi-local representation is identical to the local AdS$_4$ higher spin representation. The canonical transformation (as well as the generators) can be generalized to arbitrary higher dimensions \cite{Jevicki:2011aa}. One should also note that the $1/N$ vertices do not become local in AdS spacetime. Actually the light-cone gauge fixing of Vasiliev's full nonlinear theory has not been established yet, based on the collective map one can expect that it takes a nonlocal form.

In summary, we have demonstrated the one-to-one map between the two descriptions: the null-plane bi-locals $\Psi(x^+;x_1^-,x_2^-;x_1,x_2)$ and the higher spin fields $\mathcal{H}(x^+;x^-,x,z;\theta)$ in AdS$_4$. Both fields have same number of dimensions $1+2+2=1+3+1$, the same representation of the conformal group, and the same number of degrees of freedom. Consequently the mapping that we establish between bi-local and higher spin fields in AdS$_4$ is one-to-one and it involves the extra AdS$_4$ coordinate $z$ in a nontrivial way.

As a consequence of the above map, it follows that the wave equation in the collective picture has a map to the wave equation of higher-spin gravity in four-dimensional AdS background. This follows from the agreement of the generators (\ref{cft1}) and (\ref{ads1}) after the canonical transformation. An integral transformation between the higher-spin fields and bi-local fields was found to be
\begin{eqnarray}
&&\h(x^-,x,z,\theta)=\int dp^+ dp^x dp^z e^{i (x^- p^++x p^x+z p^z)} \cr
&&\int dp_1^+ dp_2^+ dp_1 dp_2 \delta(p_1^+ + p_2^+ - p^+)\delta(p_1+p_2-p^x) \cr
&&\delta\Bigl(p_1 \sqrt{p_2^+ / p_1^+} -p_2 \sqrt{p_1^+ / p_2^+}-p^z\Bigr) \cr
&&\delta\bigl(\theta-2\arctan\sqrt{p_2^+ / p_1^+}\bigr) \tilde{\Psi}(p_1^+,p_2^+,p_1,p_2)
\end{eqnarray}
where $\tilde{\Psi}(p_1^+,p_2^+,p_1,p_2)$ is the Fourier transform of the bi-local field $\Psi(x_1^-,x_2^-,x_1,x_2)$. 

An important check regarding the identification of the ``extra'' AdS coordinate $z$ can be seen by taking the $z \rightarrow 0$ limit. Evaluating the bi-local field at $z=0$ gives the following ``boundary'' form
\begin{equation}
\mathcal{H}(x^-,x,0,\theta) = \int dp_1^+dp_2^+e^{ix^-(p_1^++p_2^+)}
\delta(\theta-2\tan ^{-1}\sqrt{p_2^+/p_1^+})\tilde{\Psi}(p_1^+,p_2^+;x,x) \ .
\label{agreecurrent}
\end{equation}
Expanding the kernel in the above transformation into Fourier series, for a fixed even spin $s$, one finds agreement with the conformal currents of a fixed spin $s$ given by
\begin{equation}
\mathcal{O}^s=\sum_{k=0}^s\frac{(-1)^k \; \Gamma(s+\frac{1}{2})\Gamma(s+\frac{1}{2})}{k!(s-k)! \; \Gamma(s-k+\frac{1}{2})\Gamma(k+\frac{1}{2})}(\partial_-)^k \phi^a \, (\partial_-)^{s-k} \phi^a \ .
\end{equation}

As a result, in the bi-local picture one has a clear definition of the boundary $z=0$ and the notion of boundary amplitudes (boundary $S$-matrix). It is given by the relative distance between the bi-local coordinates. As such this formula does have some analogy with the short distance (renormalization group) notion \cite{Douglas:2010rc, Zayas:2013qda}, but is much more specific. Due to the construction through collective field theory, one is guaranteed to reproduce the boundary correlators in full agreement with the $O(N)$ model. The bulk/bi-local theory is nonlinear with nonlinearities governed by $1/N=G_N$ and the correlators are now reproduced in terms of Witten diagrams through higher $n$-point vertices as always in AdS duals. All this provides a nontrivial check of the collective picture and the proposal that bi-local fields provide a bulk representation of AdS$_4$ higher spin fields.

\subsection{A symmetric gauge of higher spin theory}
\label{sec:symmetric}

In the above we have described the correspondence between the bi-local large $N$ field theory and a linearized higher spin theory in one higher dimensional AdS space. The one-to-one map was established through a canonical transformation involving the phase space of the collective dipole and the phase space of the higher spin particle in AdS. As described earlier, the bi-local field theory deduced from the vector model is fully known at the nonlinear level, the nonlinearity being governed by $1/N$ as its coupling constant. On the AdS side, the remarkable construction of Vasiliev represents a closed set of nonlinear higher spin equations. In order to build a connection between these two theories (at the nonlinear level), let us consider another dramatic gauge fixing of Vasiliev theory (called the $W=0$ gauge). It is in this gauge that the higher spin fields are shown to have the same dimensionality as the {\it covariant} bi-local fields we introduced at the beginning of section \ref{sec:collective} \cite{Jevicki:2011aa}.

The full nonlinear Vasiliev theory is described in terms of three master fields: $W$, $S$ and $B$. The $W$ master field contains the higher spin degrees of freedom, while the $S$ field is purely auxiliary (mediating the interactions) and the $B$ field contains the matter degrees of freedom. In the case of AdS$_4$, the master fields are explicitly
\begin{eqnarray}
W &=& dx^\mu W_\mu (x_\mu \vert y_\alpha, \bar{y}_{\dot{\alpha}} ; z_\beta, \bar{z}_{\dot{\beta}}) \\
S &=& dz^\alpha S_\alpha (x_\mu \vert y_\alpha, \bar{y}_{\dot{\alpha}} ; z_\beta , \bar{z}_{\dot{\beta}}) + d\bar{z}^{\dot{\alpha}} \bar{S}_{\dot{\alpha}} (x_\mu \vert y_\alpha, \bar{y}_{\dot{\alpha}} ; z_\beta , \bar{z}_{\dot{\beta}}) \\
B &=& B (x_\mu \vert y_\alpha, \bar{y}_{\dot{\alpha}} ; z_\beta, \bar{z}_{\dot{\beta}})
\end{eqnarray}
where $x^\mu$ is the AdS$_4$ spacetime, $(y_\alpha, \bar{y}_{\dot{\alpha}}) \equiv Y$ are two-component complex spinors (conjugate to each other) and similarly for $(z_\alpha, \bar{z}_{\dot{\alpha}}) \equiv Z$. Therefore, the Vasiliev's equations are expressed in a $4+8$ dimensional space where the extra eight coordinates parametrize the sequence of higher spin fields with an exact higher spin gauge symmetry. For a potential exact correspondence with the $3+3$ dimensional covariant bi-local field theory one would like to demonstrate that through gauge fixing one can reduce the Vasiliev's system to a symmetrical $3+3$ dimensional base space involving a single scalar field. We will now show the existence of such a gauge and present a reduced scalar field representation of Vasiliev's theory. The steps are in part analogous to a similar gauge fixing/reduction known in self-dual Yang-Mills theory.

Following the notation of \cite{Giombi:2009wh}, the Vasiliev higher-spin equations in AdS$_4$ are given by
\begin{eqnarray}
&&d_x W+W*W=0 \label{v3eom1} \\
&&d_Z W+d_x S+[W,S]_*=0 \label{v3eom2} \\
&&d_Z S+S*S=B*(e^{i\theta_0} K dz^2 + e^{-i \theta_0} \bar{K} d\bar{z}^2) \label{v3eom3} \\
&&d_x B+W*B-B*\pi(W)=0 \label{v3eom4} \\
&&d_Z B+S*B-B*\pi(S)=0 \label{v3eom5}
\end{eqnarray}
where $K \equiv e^{z^\alpha y_\alpha}$, $\bar{K} \equiv e^{\bar{z}^{\dot{\alpha}} \bar{y}_{\dot{\alpha}}}$ are the Kleinian operators and the symmetry operator $\pi$ changes the signs of all undotted spinors
\begin{equation}
\pi(f(y,\bar{y},z,\bar{z},dz,d{\bar{z}}))=f(-y,\bar{y},-z,\bar{z},-dz,d\bar{z}) \ .
\end{equation}
This set of equations (\ref{v3eom1}-\ref{v3eom5}) is the minimal coupling Vasiliev theory where the RHS of (\ref{v3eom3}) only involves linear terms in the $B$ field. (In the non-minimal theory, there are cubic or higher order terms in $B$.) Here we also included an arbitrary phase $\theta_0$. There are two parity-conserving models for $\theta_0 = 0$ and $\theta_0 = \frac{\pi}{2}$, which are called type A and type B models in \cite{Sezgin:2003pt} respectively. In particular, the type A model corresponds to $N$-component bosons and the type B model is dual to $N$-component fermions. In general, one also has the parity-violating models ($\theta_0 \neq 0,\frac{\pi}{2}$) which are conjectured to be dual to vector models coupled with Chern-Simons gauge fields \cite{Aharony:2011jz, Giombi:2011kc}. In the following (and actually all previous sections), we will concentrate on the type A model with $\theta_0=0$ and show a connection to the bosonic vector theory.

The star product law in Vasiliev's theory is defined as
\begin{equation}
f(Y,Z)*g(Y,Z)=\int d^4 U d^4 V \, e^{u^\alpha v_\alpha+\bar{u}^{\dot{\alpha}} \bar{v}_{\dot{\alpha}}} f(Y+U,Z+U) g(Y+V,Z-V) \ ,
\end{equation}
where $U \equiv (u_\alpha,\bar{u}_{\dot{\alpha}})$, $V \equiv (v_\alpha,\bar{v}_{\dot{\alpha}})$ are the integration variables. The Vasiliev equations (\ref{v3eom1}-\ref{v3eom5}) are explicitly invariant under the higher-spin gauge transformations
\begin{eqnarray}
&&\delta W=d_x \epsilon+[W,\epsilon]_* \\
&&\delta S=d_Z \epsilon+[S,\epsilon]_* \\
&&\delta B=B*\pi(\epsilon)-\epsilon*B \ .
\end{eqnarray}
Notice that the equation (\ref{v3eom1}) is nothing but a flat connection condition for the $W$ field. At least locally one can always solve for $W$ and fix a gauge to set $W=0$. We will denote by $S'$ and $B'$ the corresponding master fields in this gauge. The reduced equations of motion from (\ref{v3eom2}, \ref{v3eom4}) then state that $S'$ and $B'$ are independent of the spacetime coordinate $x^\mu$. Explicitly, after the gauge transformation
\begin{eqnarray}
W(x \vert Y,Z)&=&g^{-1}(x \vert Y,Z)*d_x g(x \vert Y,Z) \\
S(x \vert Y,Z)&=&g^{-1}(x \vert Y,Z)*d_Z g(x \vert Y,Z)+g^{-1}(x \vert Y,Z)*S'(Y,Z)*g(x \vert Y,Z) \\
B(x \vert Y,Z)&=&g^{-1}(x \vert Y,Z)*B'(Y,Z)*\pi(g(x \vert Y,Z)) \ ,
\end{eqnarray}
the equations for $S'$ and $B'$ now take the form
\begin{eqnarray}
&&d_Z S'+S'*S'=B'*(K dz^2+\bar{K} d\bar{z}^2) \\
&&d_Z B'+S'*B'-B'*\pi(S')=0 \ .
\end{eqnarray}
Omitting the primes, we find the equations of motion in components
\begin{eqnarray}
&&\partial_\alpha S^\alpha+S_\alpha * S^\alpha=B*K \label{VSet51} \\
&&\bar{\partial}_{\dot{\alpha}}\bar{S}^{\dot{\alpha}}+\bar{S}_{\dot{\alpha}}*\bar{S}^{\dot{\alpha}} = B*\bar{K}  \label{VSet52} \\
&&\partial_\alpha\bar{S}_{\dot{\beta}}-\bar{\partial}_{\dot{\beta}}S_\alpha+[S_\alpha,\bar{S}_{\dot{\beta}}]_* =0  \label{VSet53} \\
&&\partial_\alpha B+S_\alpha*B-B*\bar{\pi}(S_\alpha)=0  \label{VSet54} \\
&&\bar{\partial}_{\dot{\alpha}}B+\bar{S}_{\dot{\alpha}}*B-B*\pi(\bar{S}_{\dot{\alpha}})=0 \ . \label{VSet55}
\end{eqnarray}
This description of Vasiliev's theory was already used by Giombi and Yin in their second evaluation of three-point correlators \cite{Giombi:2010vg}. The equations in question involve a scalar field $B$ and a vector $S$. Our goal is to gauge fix and reduce this set of equations even further and show a description in terms of a single scalar field. To this end we follow some steps known from an analogous treatment of self-dual Yang-Mills theory.

We note first, that the last two equations (\ref{VSet54}, \ref{VSet55}) for the $B$ field are not independent. Using the first two equations (\ref{VSet51}, \ref{VSet52}), one can solve for the $B$ field and check that (\ref{VSet54}, \ref{VSet55}) are satisfied (Bianchi identities). Therefore, we can totally eliminate the $B$ field and find the following five equations for the $S$ field
\begin{eqnarray}
&& F_{\alpha \dot{\beta}} \equiv \partial_\alpha\bar{S}_{\dot{\beta}}-\bar{\partial}_{\dot{\beta}}S_\alpha+[S_\alpha,\bar{S}_{\dot{\beta}}]_* =0 \ , \\
&&(\partial_\alpha S^\alpha+S_\alpha * S^\alpha)*K=(\bar{\partial}_{\dot{\alpha}}\bar{S}^{\dot{\alpha}} + \bar{S}_{\dot{\alpha}}*\bar{S}^{\dot{\alpha}})*\bar{K} \ ,
\label{reality}
\end{eqnarray}
where the last equation (\ref{reality}) is simply the reality condition for the $B$ field.

We next introduce an ansatz
\begin{eqnarray}
&& S_1=M^{-1}*\partial_1 M \ , \qquad S_2=\bar{M}^{-1}*\partial_2 \bar{M} \ , \\
&& \bar{S}_{\dot{1}}=\bar{M}^{-1}*\bar{\partial}_{\dot{1}} \bar{M} \ , \qquad \bar{S}_{\dot{2}}=M^{-1}*\bar{\partial}_{\dot{2}} M \ ,
\end{eqnarray}
which solves the equations $F_{1\dot{2}}=F_{2\dot{1}}=0$ automatically. By introducing an invariant scalar field $J=M*\bar{M}^{-1}$, the other two equations $F_{1\dot{1}}=F_{2\dot{2}}=0$ take a simple form
\begin{equation}
\bar{\partial}_{\dot{1}}(J^{-1}*\partial_1 J)=0 \ , \qquad
\partial_2(J^{-1}*\bar{\partial}_{\dot{2}} J)=0 \ . \label{Vzero2}
\end{equation}
Finally the last equation (\ref{reality}) becomes
\begin{equation}
\partial_2(J^{-1}*\partial_1 J)*K+\bar{\partial}_{\dot{1}}(J^{-1}*\bar{\partial}_{\dot{2}} J)*\bar{K}=0 \ ,
\label{Vzero3}
\end{equation}
where we have used the symmetry property $M(Y,Z)=M(-Y,-Z)$ for the bosonic case.

To summarize, we have through gauge fixing and elimination of fields reduced the Vasiliev's theory to that of a single scalar field $J$ satisfying the three equations (\ref{Vzero2}-\ref{Vzero3}). This field is hermitian in the sense that $J * \bar{J} = 1$ and it matches up with the properties of the bi-local collective field. Out of the three equations deduced above, we interpret the first two (\ref{Vzero2}) as constraints, while the third one (\ref{Vzero3}) as the equation of motion. The constraint equations appear to reduce the base space of the scalar field $J$ from $4+4$ dimensions to $3+3$ dimensions in agreement with the covariant bi-local dipole coordinate space.


\section{Higher Spin Black Hole Entropy from CFT}

Going one dimension lower, a particular aspect of the 3d higher spin gravity we are interested in is explicit black hole solutions carrying higher spin charges. Since the higher spin gauge transformations mix the metric with higher spin fields, the horizon is not a gauge-invariant concept. Focusing on the topological sector (without scalar fields) where the action can be written in terms of pure Chern-Smions, the gauge-invariant information is encoded in the holonomies. By demanding the holonomies of the new black hole solutions are the same as the BTZ black hole solution, Gutperle and Kraus \cite{Gutperle:2011kf} were able to define a higher spin black hole carrying non-zero spin-3 charge. The solution was first written down in the wormhole gauge (with smooth horizon) and an explicit gauge transformation \cite{Ammon:2011nk} was exhibited to transform the wormhole gauge to the black hole gauge with a singularity (see also \cite{Castro:2011fm}).

The holonomy conditions are also shown to be equivalent to the first law of thermodynamics \cite{Gutperle:2011kf}, from which one can calculate the black hole entropy. In the case of $SL(3,\mathbb{R}) \times SL(3,\mathbb{R})$, there exists a compact formula\footnote{More comments on this formula will be made in the Discussion section.}
\begin{equation}
S = 4 \pi \sqrt{2 \pi \, k_{CS} {\cal L}} \sqrt{1-\frac{3}{4C}} + 4 \pi \sqrt{2 \pi \, k_{CS} \bar{\cal L}} \sqrt{1-\frac{3}{4\bar{C}}}
\label{entropySL3}
\end{equation}
where $k_{CS}$ is the Chern-Simons level and
\begin{equation}
\frac{C-1}{C^{3/2}} = \sqrt{\frac{k}{32 \pi {\cal L}^3}} {\cal W} \ , \qquad
\frac{\bar{C}-1}{\bar{C}^{3/2}} = \sqrt{\frac{k}{32 \pi \bar{\cal L}^3}} \bar{\cal W} \ .
\end{equation}

Let us focus on the more general case of $hs[\lambda]$ higher spin black hole. Due to the non-local nature of the higher spin gravity, turning on the spin-3 chemical potential will lead to non-zero values of all the other higher spin fields. The black hole solution was found as a perturbative expansion of the chemical potential \cite{Kraus:2011ds} and the partition function is therefore a series expansion\footnote{There is a similar right-moving part we omit here.}
\be\label{gravres}
\log Z_{\rm BH}(\hat\tau, \alpha) = \frac{i\pi c}{12\, \hat\tau} \Bigl[ 
1 - \frac{4}{3} \frac{\alpha^2}{\hat\tau^4} + \frac{400}{27} \frac{\lambda^2-7}{\lambda^2-4} \, 
\frac{\alpha^4}{\hat\tau^8} -\frac{1600}{27}\frac{5\lambda^4-85\lambda^2+377}{(\lambda^2-4)^2}\frac{\alpha^6}{\hat{\tau}^{12}}+ \cdots \Bigr] \ ,
\ee
where we have replaced the Chern-Simons level by the central charge $c$ using (\ref{levelc}). The factor $(\lambda^2-4)$ in the denominator is due to a particular normalization for the spin-3 current. The gravity calculation is valid at large central charge $c$ and high temperature
\begin{equation}
\alpha \to 0 \ , \qquad \hat{\tau} \to 0 \ , \qquad \frac{\alpha}{\hat{\tau}^2} \quad {\rm fixed}  \ .
\end{equation}

From the dual CFT point of view, the calculation of black hole entropy amounts to calculate the partition function
\be\label{ZAdS}
Z_{\rm CFT}(\hat\tau, \alpha) = \Tr \Bigl( \hat{q}^{L_0 - \frac{c}{24}} \, y^{W_0} \Bigr)  \ , \qquad
\hat{q} = e^{2 \pi i \hat{\tau}} \ , \quad y = e^{2 \pi i \alpha} \ ,
\ee
where the trace is taken over the entire spectrum at high temperature. Here $W_0$ is the zero mode of the spin-3 current (in the $\w_\infty[\lambda]$ algebra). The standard method to calculate this character at high temperature is to do a modular transformation, in particular, the $S$-transformation
\be
\tau = -\frac{1}{\hat{\tau}} \ ,
\ee
after which the leading contribution comes from the vacuum state. Unfortunately, a general compact formula for the modular transformation of (\ref{ZAdS}) is unknown.

The bulk partition function (\ref{gravres}) is an expansion over the chemical potential $\alpha$, so it should be compared to the CFT expansion
\be\label{gravex}
Z_{\rm CFT}(\hat\tau, \alpha) = \Tr \Bigl( \hat{q}^{L_0 - \frac{c}{24}} \Bigr) + 
\frac{(2\pi i \alpha)^2}{2!}  \Tr \Bigl( (W_0)^2\, \hat{q}^{L_0 - \frac{c}{24}} \Bigr) + 
\frac{(2\pi i \alpha)^4}{4!}  \Tr \Bigl( (W_0)^4\, \hat{q}^{L_0 - \frac{c}{24}} \Bigr) + \cdots \ ,
\ee
where the odd powers of $W_0$ terms will not contribute at leading $c$. The modular transformation can be directly applied to the first term in the expansion
\be
\Tr  \Bigl( \hat{q}^{L_0 - \frac{c}{24}} \Bigr) = \sum_{r,s} 
S_{r s}  \Tr_s \Bigl( q^{L_0-\frac{c}{24}} \Bigr) 
\sim \left(\sum_r S_{r 0}\right) \, q^{-\frac{c}{24}} + \cdots \ , \qquad q = e^{2\pi i \tau}
\ee
where the sum runs over all primaries labelled by $r,s$ (with $s=0$ the vacuum representation), $S_{r s}$ is the modular $S$-matrix and the dots indicate terms exponentially suppressed at low temperature (after the modular transformation). The leading behaviour of the logarithm is then
\be
\log \Tr \Bigl( \hat{q}^{L_0 - \frac{c}{24}} \Bigr)  = - \frac{ i \pi c }{12} \tau + \cdots 
= \frac{ i \pi c }{12 \hat{\tau}} + \cdots \ ,
\ee
which reproduces precisely the $\alpha$-independent term in (\ref{gravres}). This is equivalent to the Cardy formula for the entropy. In the following we want to repeat this analysis for the other terms in the expansion (\ref{gravex}). The main technical problem we need to understand is how traces with insertion of zero modes behave under the modular $S$-transformation.

\subsection{The general strategy}

In order to explain the general strategy let us introduce a little bit of notation. For each representation of the chiral algebra (labelled by $r$), we can define a torus amplitude by 
\begin{equation}\label{Fdef}
F_r((a^1,z_1),\ldots,(a^n,z_n);\tau) = z_1^{h_1} \cdots z_n^{h_n} \, {\rm Tr}_r
\Bigl(V(a^1,z_1) \cdots V(a^n,z_n) \, q^{L_0-\frac{c}{24}}\Bigr) \ ,
\end{equation}
where $h_j$ are the conformal dimensions of the chiral fields $a^j$, i.e.\ $L_0 a^j = h_j a^j$ with $h_j\in {\mathbb N}$. In our case, the chiral fields are the higher spin currents of the $\w_\infty[\lambda]$ algebra. These functions (\ref{Fdef}) are periodic in each variable $z_j$ under the transformations
\be\label{peri}
z_j \mapsto e^{2\pi i } z_j \ , \qquad z_j \mapsto q \, z_j  \ ,
\ee
and therefore deserved to be called ``torus amplitude''. The amplitudes have a simple transformation property \cite{Zhu} under the modular group provided that all $a^j$ are Virasoro primaries
\begin{equation}\label{Ftrans}
F_r((a^1,z_1),\ldots,(a^n,z_n);\tfrac{a\tau + b }{c \tau + d}) = (c\tau +d)^{\sum_j h_j} \, 
\sum_s M_{rs}  \, F_s((a^1,z_1^{c\tau+d}),\ldots,(a^n,z_n^{c\tau+d});\tau) \ ,
\end{equation}
where $M_{rs}$ is a matrix representation of the modular group, but not dependent on $a^j$. 

We are interested in the modular transformation properties of the traces with the insertion of zero modes. Expanding the vertex operators in modes as 
\begin{equation}
V(a,z)=\sum_{m \in \mathbb{Z}} a_m \, z^{-m-h}\ ,
\end{equation}
where $h$ is the conformal dimension of $a$, the zero modes are obtained from the contour integral $\oint \frac{dz}{z} V(a,z)$, therefore we have
\be\label{zero}
\Tr_r \Bigl( a^1_0 \cdots a^n_0 \, q^{L_0-\frac{c}{24}} \Bigr) = 
\frac{1}{(2\pi i)^n} \oint \frac{dz_1}{z_1} \cdots \oint \frac{dz_n}{z_n}\, F_r((a^1,z_1),\ldots,(a^n,z_n);\tau)  \ .
\ee
Since we know the modular transformation properties of the torus amplitudes, this can now be used to deduce that of the traces with zero modes. Concentrating on the $S$-transformation which is relevant to us, we obtain from (\ref{zero}) and (\ref{Ftrans}) that
\begin{equation}
\Tr_r\Bigl(a^1_0 \cdots a^n_0 \, \hat{q}^{L_0-\frac{c}{24}} \Bigr) =
\sum_{s} S_{rs} \frac{\tau^{-n+\sum_j h_j}}{(2\pi i)^n} \!\!
\int_1^q \frac{d\tilde{z}_1}{\tilde{z}_1} \cdots \!\! \int_1^q \frac{d\tilde{z}_n}{\tilde{z}_n}\,
F_s((a^1,\tilde{z}_1),\ldots,(a^n,\tilde{z}_n);\tau) \ . \label{form1} 
\end{equation}
Here the integration contours are the result of the transformation $z\rightarrow z^\tau\equiv \tilde{z}$, which swaps the two cycles of the torus under the $S$-transformation. From now on we work on the $\tilde{z}$-plane and drop the tildes.

In the high temperature limit  $\hat\tau\rightarrow 0$ the dominant contribution to (\ref{form1}) will come from the vacuum $s=0$. Thus our calculation is reduced to doing the multiple integrals of the torus amplitudes in (\ref{form1}) with $s=0$. In order to evaluate the torus amplitudes we can use the recursion relation of \cite{Zhu} and rewrite
\begin{eqnarray}
&& F \Bigl( (a^1,z_1),(a^2,z_2),\ldots,(a^n,z_n);\tau \Bigr)
= z_2^{h_2} \cdots z_n^{h_n} \Tr \Bigl( a^1_0 \, V(a^2,z_2) \cdots  V(a^n,z_n) \, q^{L_0-\frac{c}{24}} \Bigr) \cr
&& \qquad + \sum_{j=2}^{n} \, \sum_{m \in \mathbb{N}_0} \, {\cal P}_{m+1}\left(\frac{z_j}{z_1},q\right) \, 
F\Bigl( (a^2,z_2),\ldots, (a^1[m]a^j,z_j),\ldots , (a^n,z_n);\tau \Bigr) \ ,
\label{recur}
\end{eqnarray}
where we dropped the index $s=0$ for the torus amplitudes and the traces. The Weierstrass functions ${\cal P}$ are defined by a power series 
\be\label{defp}
{\cal P}_k(x,q) = \frac{(2\pi i)^k}{(k-1)!} \sum_{n\neq 0} \left(
\frac{n^{k-1}x^n}{1-q^n} \right) \ ,
\ee
which satisfy an important recursion relation
\be\label{Prec}
x \frac{d}{dx} {\cal P}_k(x,q) = \frac{k}{2\pi i} {\cal P}_{k+1}(x,q) \ .
\ee
The bracketed modes in the second line of (\ref{recur}) take the form
\be\label{abradef}
a[m] = (2\pi i)^{-m-1}\sum_{i=m}^\infty c(h_a,i,m)a_{-h_a + 1 + i} \ ,
\ee
where the coefficients $c(h,i,m)$ are found via the expansion
\begin{equation}
(\log (1+z))^s \, (1+z)^{h-1} = \sum_{i \ge s} c(h,i,s) \, z^i \ .
\end{equation}

With the help of this recursion relation we can then evaluate the amplitudes in (\ref{form1}) explicitly (and recursively). The recursion formula turns the amplitude into a sum of products of Weierstrass functions; then the contour integrals are straightforward because ${\cal P}(z)/z$ is a total derivative. This computation will be explained in detail for the two-point case in the following subsection (and more details for the higher points can be found in \cite{Gaberdiel:2012yb}). Fortunately, we do not need to evaluate the exact answer, but only the leading order contribution as $c \rightarrow \infty$.

\subsection{Two-point calculation}

The quadratic correction in the expansion (\ref{gravex}) after using (\ref{form1}) is
\be
Z^{(2)} \equiv \frac{(2 \pi i\alpha)^2}{2!}\Tr\bigl( (W_0)^2 \hat{q}^{L_0-\frac{c}{24}} \bigr)
\approx \frac{\alpha^2\tau^4}{2} \, \int_1^q \frac{dz_1}{z_1} \int_1^q \frac{dz_2}{z_2} \, F\big((W,z_1), (W,z_2); \tau\big) \ ,
\ee
where the symbol `$\approx$' means that we have dropped terms that do not contribute to $\log Z$ at leading order. Applying the recursion relation (\ref{recur}), we have
\be
F\big((W,z_1), (W,z_2); \tau\big) = z_2^3 \Tr\bigl( W_0 W(z_2)\, q^{L_0-\frac{c}{24}} \bigr)
+ {\cal P}_{m+1}\left( \frac{z_2}{z_1} \right) F\big((W[m]W, z_2); \tau\big) \ .
\ee
Here, and in the following, we shall always imply a sum over the Weierstrass index $m \geq 0$. In the first term, only the zero mode of $W(z_2)$ survives in the trace, but $\Tr_0 \bigl((W_0)^2 q^{L_0-c/24}\bigr)$ is exponentially suppressed. In the second term, the vacuum contribution to the trace dominates, therefore the quadratic correction becomes
\be
Z^{(2)} \approx \half q^{-c/24} \alpha^2 \tau^4 \langle W[m]W_{-3} \rangle 
\, \int_1^q \frac{dz_1}{z_1} \int_1^q \frac{dz_2}{z_2} {\cal P}_{m+1}\left( \frac{z_2}{z_1} \right) \ ,
\ee
where $\langle \cdot \rangle$ is the vacuum expectation value (on the sphere). According to (\ref{Prec}), integrating a Weierstrass function reduces its index by 1. Since only ${\cal P}_1$ is not periodic (see appendix A of \cite{Gaberdiel:2012yb} for more details), this implies the integrals
\be
\int_1^q \frac{dz_2}{z_2}\, {\cal P}_2\left( \frac{z_1}{z_2} \right) = (2\pi i)^2 \  , \qquad 
\int_1^q \frac{dz_2}{z_2}\, {\cal P}_{m+1}\left( \frac{z_1}{z_2} \right) = 0 \quad (m>1) \ .
\ee
Therefore the quadratic correction is now simplified to
\be
Z^{(2)}\approx \half q^{-c/24}(2\pi i)^3 \alpha^2 \tau^5 \langle W[1]W_{-3}\rangle \ .
\ee
From the definition of the bracketed mode (\ref{abradef}), we have
\be
W[1] = (2\pi i)^{-2}\left(W_{-1} + \frac{3}{2}W_0 + \frac{1}{3}W_{1} - \frac{1}{12}W_{2} + \frac{1}{30}W_{3} + \cdots \right) \ .
\ee
Since we are only interested in the vacuum expectation value, the only surviving term is then $\langle W[1]W_{-3}\rangle = (2\pi i)^{-2}\tfrac{1}{30} \langle W_3 W_{-3} \rangle$. Plugging in the commutators of $\w_\infty[\lambda]$ (see appendix D of \cite{Gaberdiel:2012yb}), the final answer is 
\be
Z^{(2)} = 2 \pi i c\, \frac{\alpha^2\tau^5}{18} \,q^{-c/24}\ ,
\ee
in precise agreement with the leading correction to the gravitational partition function (\ref{gravres}). 

This calculation can be extended to higher points (the four-point and six-point calculations were presented in \cite{Gaberdiel:2012yb}) showing exact agreement with the gravity result (\ref{gravres}). A genuine difference between the two-point calculation and the higher-point computation is that the nonlinear terms of the $\w_\infty[\lambda]$ algebra (which are proportional to $1/c$) plays an essential role to match the gravity result. The gives a detailed and different check of the $\w_\infty[\lambda]$ algebra as the symmetry of the boundary CFT.


\section{Extensions and Discussion}

Returning to the entropy formula (\ref{entropySL3}) for $SL(3)$ higher spin black holes, it can also be evaluated using the on-shell Chern-Simons action \cite{Banados:2012ue} and the conical singularity approach \cite{Kraus:2013esi}. These methods are shown to satisfy the same integrability conditions or the first law of thermodynamics  as in \cite{Gutperle:2011kf}; thus they give consistent results. On the other hand, by transforming the frame-like formulation of higher spin gravity to the metric-like formulation \cite{Campoleoni:2012hp}, the black hole entropy can also be calculated using the Wald's formula. However, the result of \cite{Campoleoni:2012hp} differs from (\ref{entropySL3}) by an additional factor of ($1-\frac{3}{2C}$). A generalized Bekenstein-Hawking formula was further derived in \cite{Perez:2013xi} showing agreement with \cite{Campoleoni:2012hp}. The two seemingly different results was clarified in \cite{deBoer:2013gz} (from the bulk side) as different boundary terms added to the pure Chern-Simons action. This leads to different ``definitions'' of the energy and therefore different expressions for the entropy. It is also interesting and important to understand this subtle point regarding the CFT computation.

The pure Chern-Simons theory in 3d is topological with no propagating degrees of freedom. The proposal of Gaberdiel and Gopakumar \cite{Gaberdiel:2010pz} does include the scalar fields. Therefore, one is also interested in studying the scalar propagation in higher spin gravity, especially on the backgrounds of higher spin black holes. At linearized level, the master field $C$ containing the matter degrees of freedom satisfies the equation of motion
\begin{equation}
d C + A_0 * C - C * \bar{A}_0 = 0 \ ,
\end{equation}
where $(A_0,\bar{A}_0)$ are the higher spin background gauge fields. Explicit bulk-boundary propagators can be calculated for the $hs[\lambda]$ black hole \cite{Kraus:2012uf} and the boundary two-point function of the scalars can be deduced by taking certain limits. They are seen to match exactly a direct CFT calculation \cite{GJP} (again using the ${\cal W}_\infty[\lambda]$ algebra). This gives further evidence of the AdS$_3$/CFT$_2$ correspondence.

There are many other interesting generalizations and aspects of the Higher Spin/Vector Model correspondence, some of which are for example: higher spin black hole solutions with spin-4 chemical potential \cite{Tan:2011tj, Chen:2012pc}; the conical defect solutions and their mapping to the light states in the dual CFT \cite{Castro:2011iw, Perlmutter:2012ds, Campoleoni:2013lma}; the minimal model holography for the other cosets \cite{Ahn:2011pv, Gaberdiel:2011nt, Candu:2012ne}; the supersymmetric generalization of various AdS$_4$/CFT$_3$ and AdS$_3$/CFT$_2$ dualities \cite{Engquist:2002vr, Leigh:2003gk, Creutzig:2011fe, Candu:2012jq, Candu:2012tr}; the higher spin theory in de Sitter spacetime and its duality to $sp(2N)$ vector model \cite{Anninos:2011ui, Das:2012dt, Anninos:2012ft}; the derivation of higher spin theory from certain limits of string theory \cite{Sundborg:2000wp, Polyakov:2009pk, Sagnotti:2010at, Chang:2012kt}, etc. The reader can refer to some recent review papers \cite{Vasiliev:2012vf, Gaberdiel:2012uj, Giombi:2012ms, Ammon:2012wc, Sezgin:2012ag, Jevicki:2012fh} for more details and references.

\acknowledgments

The author would like to thank the local organizers for their hospitality during the workshop. The first part of review is based on the collaboration with Antal Jevicki, Robert de Mello Koch, Jo\~{a}o Rodrigues and Qibin Ye; while the second part of the review is based on the work done with Matthias Gaberdiel and Tom Hartman. It is also a great pleasure to thank Sumit Das, Soo-Jong Rey, Constantin Candu, Eric Perlmutter, Cristian Vergu and all the participants of the workshop for useful discussions.


\end{document}